\documentclass[universe,article,accept,pdftex,moreauthors]{Definitions/mdpi}

\firstpage{1} 
\pubvolume{1}
\issuenum{1}
\articlenumber{0}
\pubyear{2024}
\copyrightyear{2024}
\externaleditor{Academic Editor: }
\datereceived{ } 
\daterevised{ } 
\dateaccepted{ } 
\datepublished{ } 
\hreflink{https://doi.org/} 

\usepackage{amsmath} 
\newcommand{\angstrom}{{\rm \mathring A}}
\usepackage{aas_macros}

\Title{The Host Galaxy Fluxes of Active Galaxy Nuclei Are Generally Overestimated by the Flux Variation Gradient Method}

\TitleCitation{The Host Galaxy Fluxes of Active Galaxy Nuclei Are Generally Overestimated by the Flux Variation Gradient Method}

\Author{\hl{Min-Xuan Cai} 
$^{1,2,}$\hl{*} 
\orcidA{}, Zhen Wan $^{1,2}$\orcidB{}, \hl{Zhen-Yi Cai} $^{1,2,}$\hl{*}\orcidC{}, \hl{Lu-Lu Fan} $^{1,2,3}$\orcidD{} and \hl{Jun-Xian Wang} $^{1,2}$\orcidE{}}

\AuthorNames{Min-Xuan Cai, Zhen Wan, Zhen-Yi Cai, Lu-Lu Fan  and Jun-Xian Wang}

\AuthorCitation{\hl{Cai, M.-X.; Wan, Z.; Cai, Z.-Y.; Fan, L.-L.; Wang, J.-X.} 
}

\address{%
$^{1}$ \quad CAS Key Laboratory for Research in Galaxies and Cosmology, Department of Astronomy,
University of Science and Technology of China, Hefei 230026, China; \hl{zhen\_wan@ustc.edu.cn (Z.W.); llfan@ustc.edu.cn~(L.-L.F.); jxw@ustc.edu.cn (J.-X.W.)} 
\\
$^{2}$ \quad School of Astronomy and Space Science, University of Science and Technology of China, Hefei 230026, China\\
$^{3}$ \quad Deep Space Exploration Laboratory, Hefei 230088, China}

\corres{Correspondence: caimx@mail.ustc.edu.cn (M.-X.C.); zcai@ustc.edu.cn (Z.-Y.C.)}

\abstract{In terms of the variable nature of normal active galaxy nuclei (AGN) and luminous quasars, a so-called flux variation gradient (FVG) method has been widely utilized to estimate the underlying non-variable host galaxy fluxes. The FVG method assumes an invariable AGN color, but this assumption has been questioned by the intrinsic color variation of quasars and local Seyfert galaxies. Here, using an up-to-date thermal fluctuation model to simulate multi-wavelength AGN variability, we theoretically demonstrate that the FVG method generally overestimates the host galaxy flux; that is, it is more significant for brighter AGN/quasars. Furthermore, we observationally confirm that the FVG method indeed overestimates the host galaxy flux by comparing it to that estimated through other independent methods. We thus caution that applying the FVG method should be performed carefully in the era of time-domain astronomy.}


\begin{document}


\section{Introduction} \label{sec:intro}

\hl{Residing} 
 in a very tiny center region of a galaxy, energetic active galactic nuclei (AGN), including luminous quasars and low-luminosity analogs, behave as a point-like source, superimposed on its host galaxy, which can have diverse morphologies~\citep{2009ApJ...691..705G}. For luminous quasars, the optical AGN emission takes over from the host galaxy, while the host galaxy contribution is comparable to, or even larger than, that of the AGN at longer infrared wavelengths and in low-luminosity Seyfert galaxies~\citep{Fausnaugh2016ApJ...821...56F}. Accurately separating the AGN emission and its host galaxy contribution has significant implications for both the AGN and host galaxy studies, such as photometric reverberation mapping~\citep{2019NatAs...3..251C,2019MNRAS.490.3936P,2023MNRAS.522.2002P,Ma2023ApJ...949...22M}, the broad-line region size versus luminosity relation~\citep{2013ApJ...767..149B}, and the host galaxy properties~\citep{2016ApJ...826...62H}. Thus, it is essential to disentangle the respective contributions from the AGN and host galaxy in order to properly investigate the 
corresponding physical properties.

Two traditional methods have been proposed to disentangle fluxes of the AGN and host galaxy. One of them involves fitting galaxy and AGN templates to the observed spectrum, e.g., \citep{2015A&A...575A..22M, 2016ApJ...826...62H, 2019ApJ...886..150M}, while the other one involves modeling the host galaxy profile using a high-resolution image, e.g., \citep{1993MNRAS.263..655K, 2002A&A...384..780B, 2009ApJ...697..160B}. The former method requires high-resolution spectroscopic data, while the latter is largely restricted to local Seyfert galaxies whose host galaxies are spatially resolvable. And, both of them are time consuming and technically complex. 

In the era of time-domain astronomy, there is a simple method proposed as the flux variation gradient (FVG) method~\citep{1992MNRAS.257..659W}, which estimates the host galaxy contribution on the basis of multi-band AGN light curves. The FVG method has been widely employed in many studies to decompose fluxes of the AGN and its host galaxy for some specific objects, typically local AGN, e.g., \citep{2015A&A...581A..93R,2019NatAs...3..251C,2019MNRAS.490.3936P,2020AJ....159..259S, 2023A&A...672A.132F}, and utilized for other applications~\citep{2015A&A...578A..98P,2022MNRAS.516.2876M,2023MNRAS.518..418H}. More recently, an upgraded probabilistic FVG method has been proposed~\citep{2022A&A...657A.126G, 2023MNRAS.525.4524G}.

Originally proposed by~\citet{1992MNRAS.257..659W}, the FVG method is based on the observed linear relation of fluxes between two bands~\citep{1981AcA....31..293C}, which decomposes the total (host galaxy + AGN) flux into a constant (red) host galaxy component and a variable (blue but invariable color) AGN component in order to explain the bluer-when-brighter phenomenon observed in active galaxies. However, there are also some works indicating that the FVG method may not be as accurate as previously thought. Notably,~\citet{2011A&A...535A..73H} discovered that the host galaxy fluxes obtained using the FVG method are twice as high as those obtained through image decomposition. Later,~\citet{2014ApJ...792...54S} unveiled the intrinsic timescale-dependent bluer-when-brighter feature of quasars, which challenges the basic assumption of the FVG method. Thus,~\citet{2014ApJ...792...54S} suggested that the FVG method can not be applicable for quasars. 
Moreover,~\citet{2019ApJ...886..150M} found that when applying to several quasars, the FVG method tends to overestimate the host galaxy flux and has larger uncertainties when compared to the spectral and image decomposition methods.

In this work, we explore the accuracy of the FVG method in recovering the host galaxy flux. First, we demonstrate in Section~\ref{sec:simulation} that the FVG method tends to overestimate the host galaxy flux using multi-band AGN light curves simulated by an up-to-date thermal fluctuation model. In Section~\ref{sec:Applications}, by comparing the host galaxy fluxes obtained through the FVG and image decomposing methods, we confirm that the FVG method indeed results in a higher host galaxy contribution. Section~\ref{sec:discussion} discusses the reliability and the potential issues of the FVG method. Finally, conclusions are presented in Section~\ref{sec:conclusion}.

\section{Theoretical Predictions} \label{sec:simulation}

\subsection{Simulating AGN Light Curves}

The FVG method is based on the assumption that the intrinsic AGN color remains unchanged. However, the discovery of the intrinsic timescale-dependent bluer-when-brighter feature of quasars by~\citet{2014ApJ...792...54S} has challenged this assumption. Later on,~\citet{Cai2016} proposed a revised inhomogeneous thermal fluctuation model based on~Dexter and Agol~\cite{2011ApJ...727L..24D} to explain the timescale-dependent bluer-when-brighter feature observed by~\citet{2014ApJ...792...54S}. Subsequent upgrades on the thermal fluctuation model further account for the lag as well as the correlation among multiple wavebands across optical to X-ray~\citep{Cai2018,Cai2020}. See Section~\ref{sect:tfm} for a discussion on improving the thermal fluctuation model.

Here, we use the up-to-date thermal fluctuation model~\citep{Cai2018,Cai2020} to simulate multi-wavelength emissions for AGN, akin to NGC 5548. The simulated AGN light curves cover a baseline of $\sim$180 days with time steps of 0.04 day (see Cai et~al.~\cite{Cai2020} for details). The baseline is also typical for most AGN observed by ground-based telescopes per year. We consider light curves in nine bands, including $BVRI$ of Johnson/Cousins (JC) and $ugriz$ of Sloan Digital Sky Survey (SDSS). In total, 200 independent simulations are performed, and we confirm that the number of simulations is large enough for examining the FVG method.
In this work, we primarily focus on the SDSS $u$ band and $g$ band light curves, which are widely analyzed by previous FVG studies.

\subsection{Adding Host Galaxy Contribution}

To construct the total light curves including both the AGN and host galaxy contributions, we define a ratio of the host galaxy flux to the total flux in the $g$ band, $\alpha_g$ (hereafter, the host galaxy flux fraction), and then derive the host galaxy fluxes in the $u$ band by assuming a specific spectral energy distribution (SED) for the host galaxy. 

Specifically, we define $\alpha_g \equiv f^{\rm H}_g / \langle f_g \rangle$ and $f_{g}(t) = f^{\rm A}_{g}(t) + f^{\rm H}_{g}$, where $f^{\rm H}_{g}$ is the invariable $g$ band host galaxy flux, $f^{\rm A}_{g}(t)$ is the simulated variable $g$ band AGN flux whose mean is $\langle f^{\rm A}_{g} \rangle$, and $f_{g}(t)$ is the variable $g$ band total flux whose mean is $\langle f_{g} \rangle$. Equivalently, $f^{\rm H}_{g} = \langle f^{\rm A}_{g} \rangle \alpha_g / (1 - \alpha_g)$.
From the $g$ band to $u$ band, we have the $u$ band host galaxy flux as
\begin{equation}\label{eq:host_g_to_u}
    f^{\rm H}_u = \frac{\int_u S(\lambda) T_u(\lambda) d\lambda / \int_u T_u(\lambda) d\lambda}{\int_g S(\lambda) T_g(\lambda) d\lambda / \int_g T_g(\lambda) d\lambda} f^{\rm H}_g,
\end{equation}
where $S(\lambda)$ is the SED assumed for the host galaxy and $T_u$ (or $T_g$) is the $u$ band (or $g$ band) transmission curve. Finally, the variable $u$ band total flux $f_{u}(t) = f^{\rm A}_{u}(t) + f^{\rm H}_u$, where $f^{\rm A}_{u}(t)$ is the simulated variable $u$ band AGN flux.
Hereafter, $f_{x}(t)$, $f^{\rm A}_{x}(t)$, and $f^{\rm H}_{x}$ indicate the total, AGN, and host galaxy fluxes in the $x$ band, respectively. 

\begin{figure}[H]

    \includegraphics[width=0.8\linewidth]{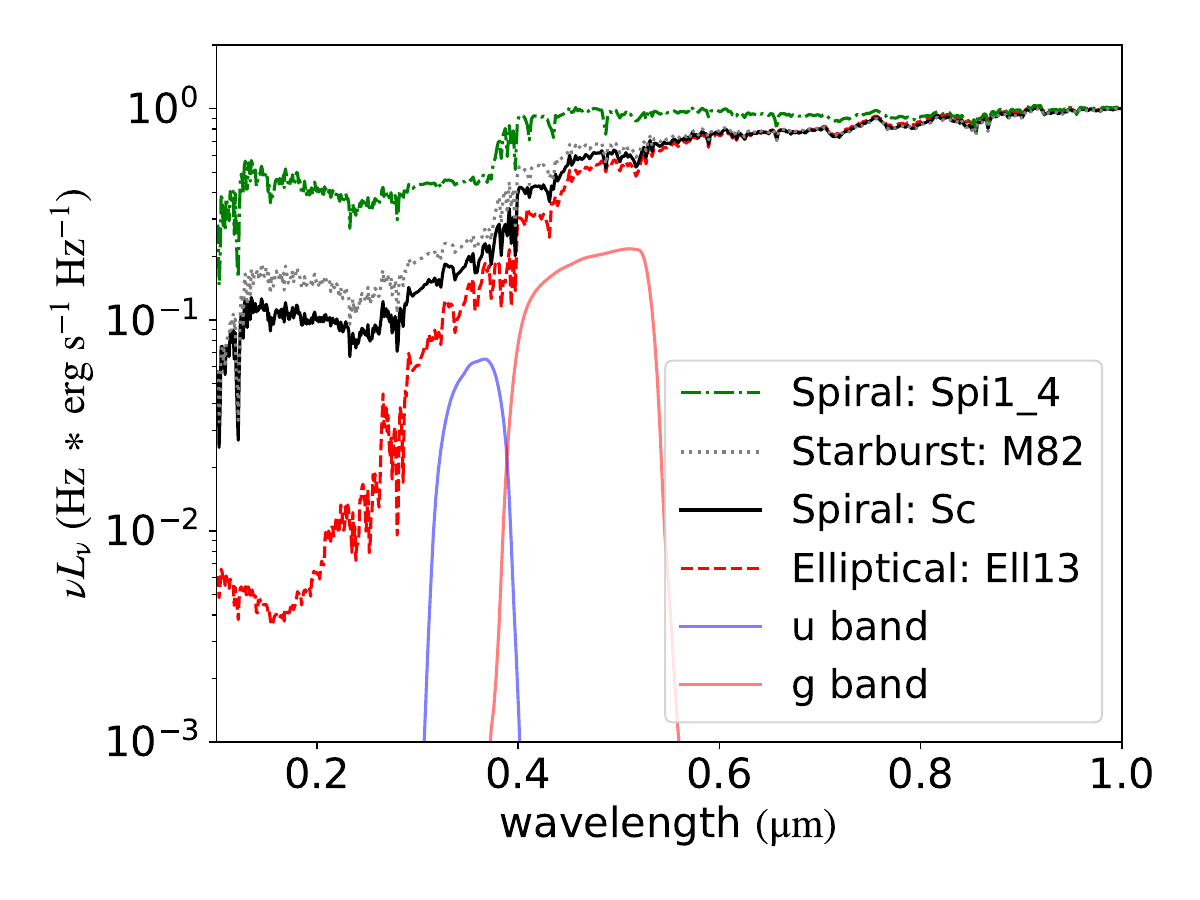}
    \caption{Illustrated are four SEDs of distinct host galaxies ranging from the early elliptical galaxy to the late spiral galaxy as well as starburst types with the SED types of Ell13, Sc, M82, and Spi1\_4 from~\citet{SED_swire}. These SEDs are normalized at 1 $\upmu$m to highlight the difference at shorter wavelengths. Transmission curves of the SDSS $u$ and $g$ bands are shown for comparison.}
    \label{fig:SED_compare}
\end{figure}

To exhaustively explore how well the host galaxy flux can be reproduced using the FVG method, given different host galaxy contributions and galaxy types, we consider $\alpha_g$ from 0.1 to 0.9 in steps of 0.1 and galaxy types ranging from the early elliptical galaxy to the late spiral galaxy as well as the starburst galaxy. Four SED templates representing Ell13, Sc, Spi1\_4, and M82 from Polletta et~al.~\cite{SED_swire} are illustrated in Figure~\ref{fig:SED_compare}. All other star-forming and starburst SEDs from Polletta et~al.~\cite{SED_swire} are enclosed between the lowest Ell13 SED and the highest Spi1\_4 SED, while the Sc SED is almost in the middle. Figure~\ref{fig:SED_compare} also shows the SED from the starburst galaxy, M82, which is very similar to the Sc SED. 

The similarity between the Sc and M82 SEDs is the result of a competition between star formation and dust extinction: starburst galaxies have a larger star formation rate but also more dust extinction than spiral galaxies. 
In the following, we would only discuss the FVG method assuming the Ell13, Sc, and Spi1\_4 SEDs since the Sc SED is closer to the middle than the M82 one. Considering any other SEDs does not alter our conclusions.

\subsection{Simulating Observations}\label{sec:Mocking}

To further investigate how the real observational conditions, such as the observed duration ($T$), the cadence ($\Delta t$), and the signal-to-noise ratio (SNR or ${\rm S/N}$), affect the accuracy of the FVG method, we simulated a series of observations with $T = 45$, 90, and 180 days; $\Delta t = 1$, 3, and 5 days; and ${\rm S/N}= 25$, 50, and 100. We take $\{T, \Delta t, {\rm S/N}\} = \{90, 3, 50\}$ as the fiducial simulation given a series of $\alpha_g$ and discuss the effect of changing any condition on the output of applying the FVG method.

\begin{figure}[H]

    \includegraphics[width=1.0\textwidth]{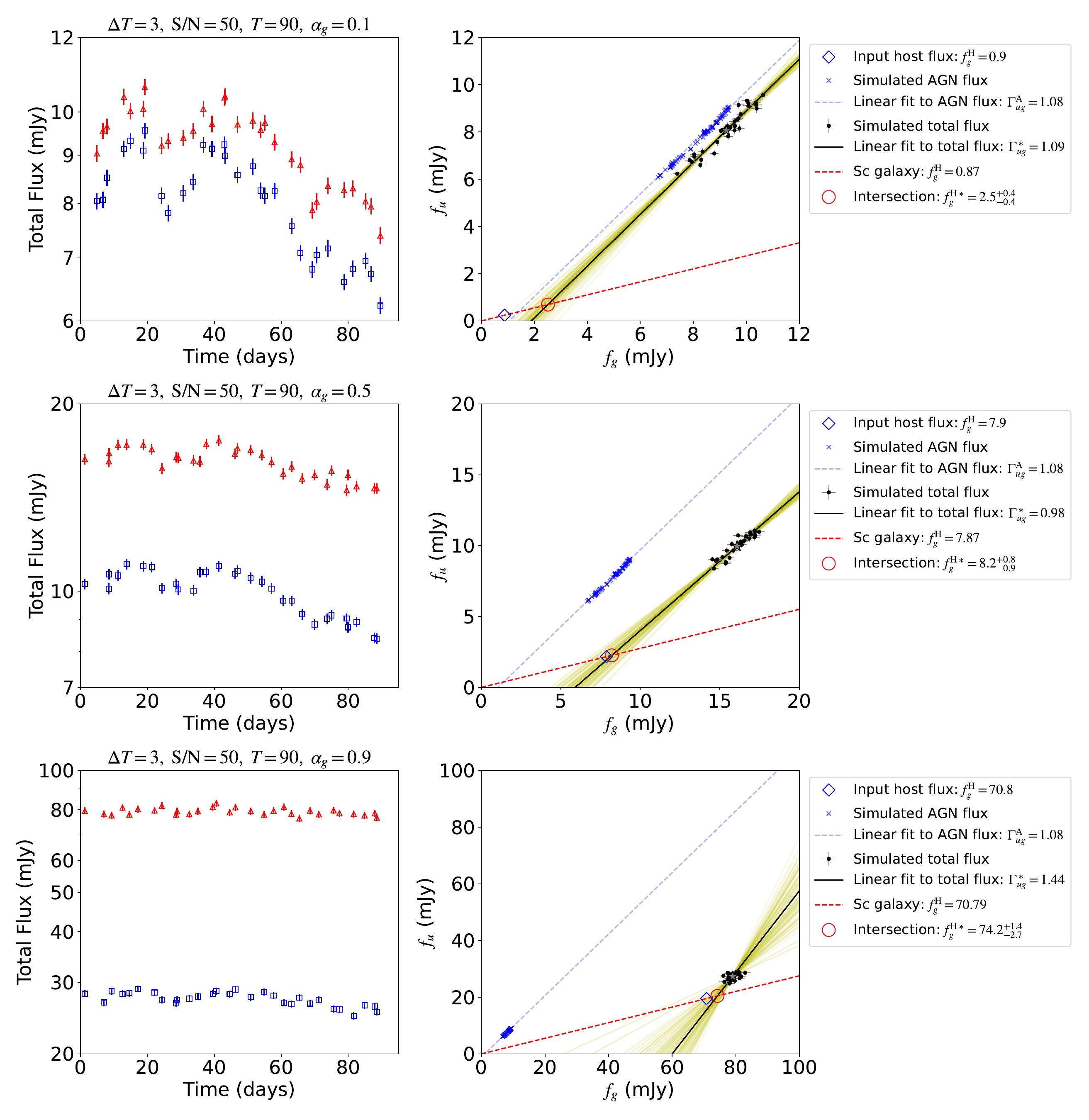}
    \caption{For our fiducial simulated observation (i.e., $T = 90$ days, $\Delta T = 3$ days, and $S/N = 50$), applications of the FVG method are illustrated using the simulated $u$ and $g$ band light curves (\textbf{left} panels; blue open squares represent the $u$ band and red open triangles represent the $g$ band) for different $g$ band host galaxy flux fractions: $\alpha_g = 0.1$ (\textbf{top} panels), 0.5 (\textbf{middle} panels), and 0.9 (\textbf{bottom} panels). 
    In each right panel, the simulated pure AGN fluxes in the two bands (blue crosses) are linearly fit by a light-blue dashed line, while the simulated total fluxes in two bands (black filled circles with error bars) are linearly fit by a black solid line. In each right panel, we randomly selected 100 sets of parameters from our MCMC sampling and plotted the corresponding lines in yellow.
    The intersection between the black solid line and the red dashed line (i.e., the ratio of $u$ band to $g$ band fluxes inferred by assuming the Sc SED for the host galaxy) indicates the retrieved host galaxy flux (red open circle), which is larger than the input host galaxy flux (blue open diamond). In all cases, values of the retrieved $g$ band host galaxy flux, $f^{\rm H*}_g$, are found to be larger than those of the input $g$ band host galaxy flux, $f^{\rm H}_g$.}
    \label{fig:FVG_sketch}
\end{figure}

For our fiducial simulation, the $u$ and $g$ band total (AGN + host galaxy) light curves are illustrated in the left panels of Figure~\ref{fig:FVG_sketch} for $\alpha_g = 0.1$, 0.5, and 0.9 from top to bottom, respectively. Here, the simulated light curves are linearly interpolated from the ideal 0.04 days to the desired $\Delta t$, and then all fluxes are randomly offset by a Gaussian distribution with zero mean and $f_x/({\rm S/N})$ as the standard deviation.

\subsection{Retrieving the Host Galaxy Flux with the FVG Method}\label{sec:sim_result}

Now we use the FVG method to retrieve the host galaxy fluxes from the simulated light curves. 
Denote the total fluxes in the $g$ and $u$ bands as $\{..., (f_{g, i}, f_{u, i}), ...\}$ observed at times $\{..., t_i, ...\}$ with 1$\sigma$ measurement errors $\{..., (\sigma_{g, i}, \sigma_{u, i}), ...\}$ (the left panels of Figure~\ref{fig:FVG_sketch}).
The likelihood of observing $(f_{g, i}, f_{u, i})$ given a straight line, $y = kx + b$ (the black lines in the right panels of Figure~\ref{fig:FVG_sketch}), is determined according to a specific point on a two-dimensional Gaussian surface. The two-dimensional Gaussian surface is centered at $(f_{g, i}, f_{u, i})$ with variances $(\sigma_{g, i}, \sigma_{u, i})$, while the specific point, $(f^{\rm T}_{g, i}, f^{\rm T}_{u, i})$, is the point of tangency of the straight line to an ellipse with an eccentricity of $\sqrt{1-(\sigma_{u,i} / \sigma_{g,i})^2}$ on the two-dimensional Gaussian surface: $f^{\rm T}_{u,i} = k f^{\rm T}_{g,i} + b$ and $f^{\rm T}_{g,i} = \frac{k (f_{u,i} - b) \sigma_{g,i}^2+ f_{g,i} \sigma_{u,i}^2}{k^2 \sigma_{g,i}^2 + \sigma_{u,i}^2}$.
Specifically, the likelihood of observing $(f_{g, i}, f_{u, i})$ is given by
\begin{equation}
    \label{eq:mcmc_likelihood_function}
    \ln L_i[(f_{g, i}, f_{u, i})|k, b] = -\frac{1}{2} \left[ \left( \frac{f^{\rm T}_{g,i} - f_{g,i}}{\sigma_{g,i}} \right)^2 + \left( \frac{f^{\rm T}_{u,i} - f_{u,i}}{\sigma_{u,i}} \right)^2 \right],
\end{equation}
and thus the likelihood of observing $\{..., (f_{g, i}, f_{u, i}), ...\}$ is $\ln L = \sum_i \ln L_i$, which is maximized using the Markov Chain Monte Carlo (MCMC) sampling by emcee, a python module which implemented this algorithm~\citep{2013PASP..125..306F}. Thus, the best-fit value and the associated $1 \sigma$ uncertainty for the free parameter $k$ (and $b$) are estimated according to the median and 16th to 84th percentiles of the corresponding one-dimensional marginalized posterior probability distribution, respectively. Finally, the retrieved host galaxy fluxes, $(f^{\rm H*}_g, f^{\rm H*}_u)$, are the intersection point (the red circles in the right panels of Figure~\ref{fig:FVG_sketch}) of the straight line and a line (the red dashed lines in the right panels of Figure~\ref{fig:FVG_sketch}) passing through the origin with a slope, $\Gamma^{\rm H}$, determined by the assumed host galaxy SED.

In practice, we replace $y = k x + b$ with $y = \Gamma^{\ast}_{ug}x + (\Gamma^{\rm H} - \Gamma^{\ast}_{ug})f^{\rm H\ast}_g$, where $k = \Gamma^{\ast}_{ug}$ and $b = (\Gamma^{\rm H} - \Gamma^{\ast}_{ug})f^{\rm H\ast}_g$, such that we can directly sample the $g$ band host galaxy flux, $f^{\rm H\ast}_g$. Uniform priors on $\Gamma^{\ast}_{ug}$ and $f^{\rm H\ast}_g$ are assumed. The dynamical range of $f^{\rm H\ast}_g$ is limited to $[0, \min\{f_{g,i}\}]$ since the retrieved host galaxy flux should always be greater than zero and should not exceed the total flux.
The right panels of Figure~\ref{fig:FVG_sketch} illustrate how the FVG method works for $\alpha_g = 0.1$, 0.5, and 0.9 from top to bottom, respectively.

\subsection{The FVG Method Generally Results in Biased Host Galaxy Fluxes}

\begin{figure}[H]

    \begin{minipage}{0.45\linewidth}
        \includegraphics[width=1.0\linewidth]{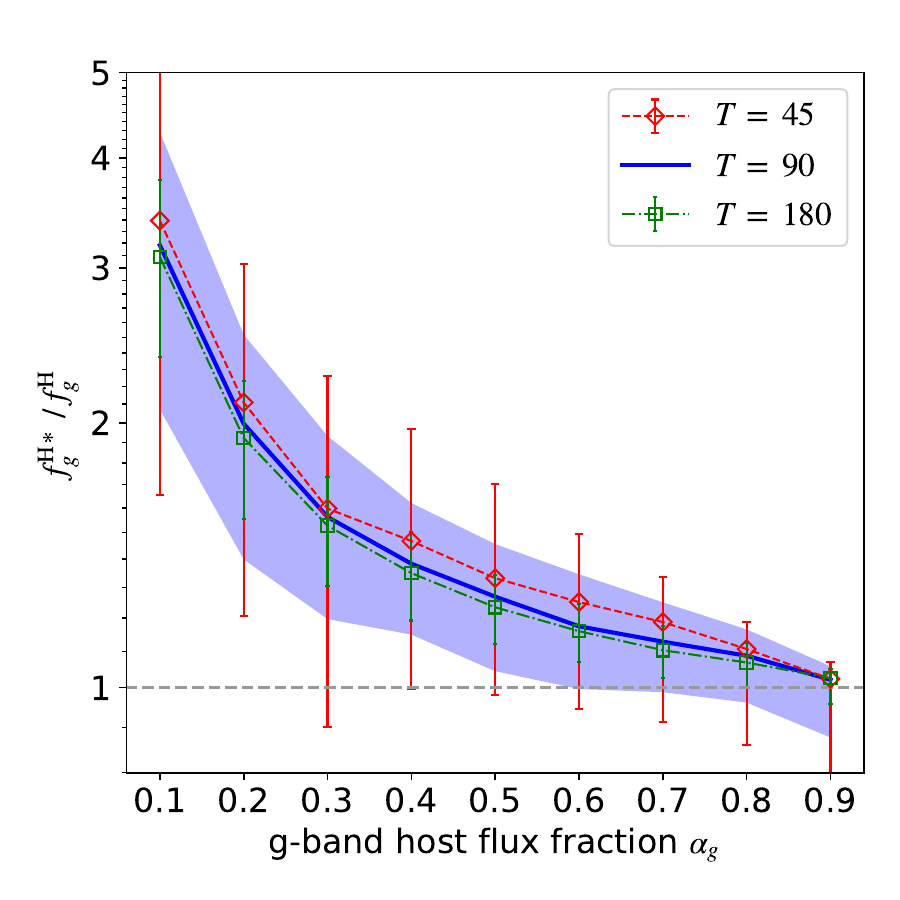}
    \end{minipage}
    \begin{minipage}{0.45\linewidth}
       \includegraphics[width=1.0\linewidth]{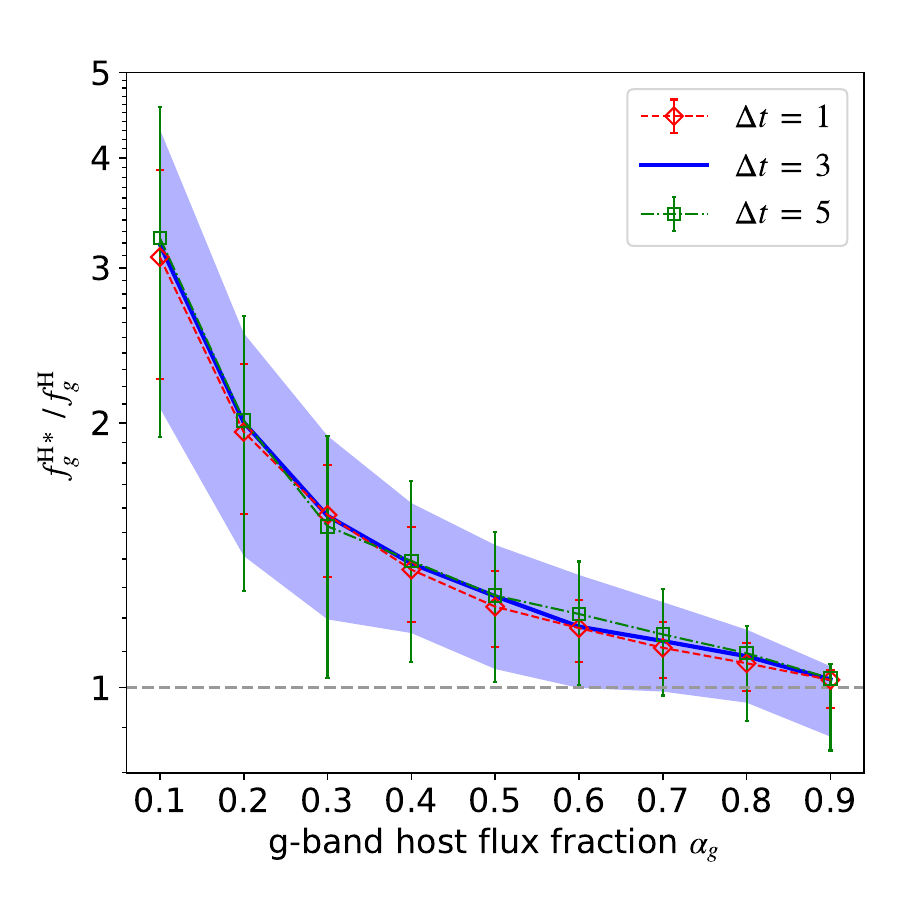}
    \end{minipage}\\
    \begin{minipage}{0.45\linewidth}
        \includegraphics[width=1.0\linewidth]{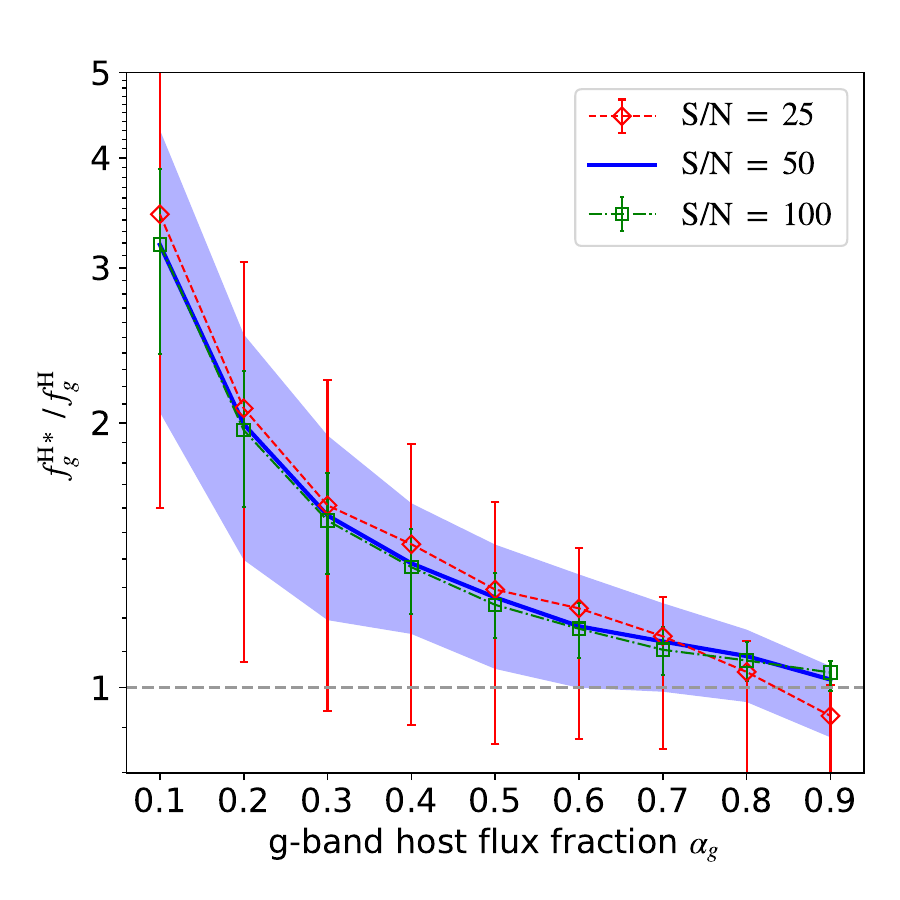}
    \end{minipage}
    \begin{minipage}{0.45\linewidth}
        \includegraphics[width=1.0\linewidth]{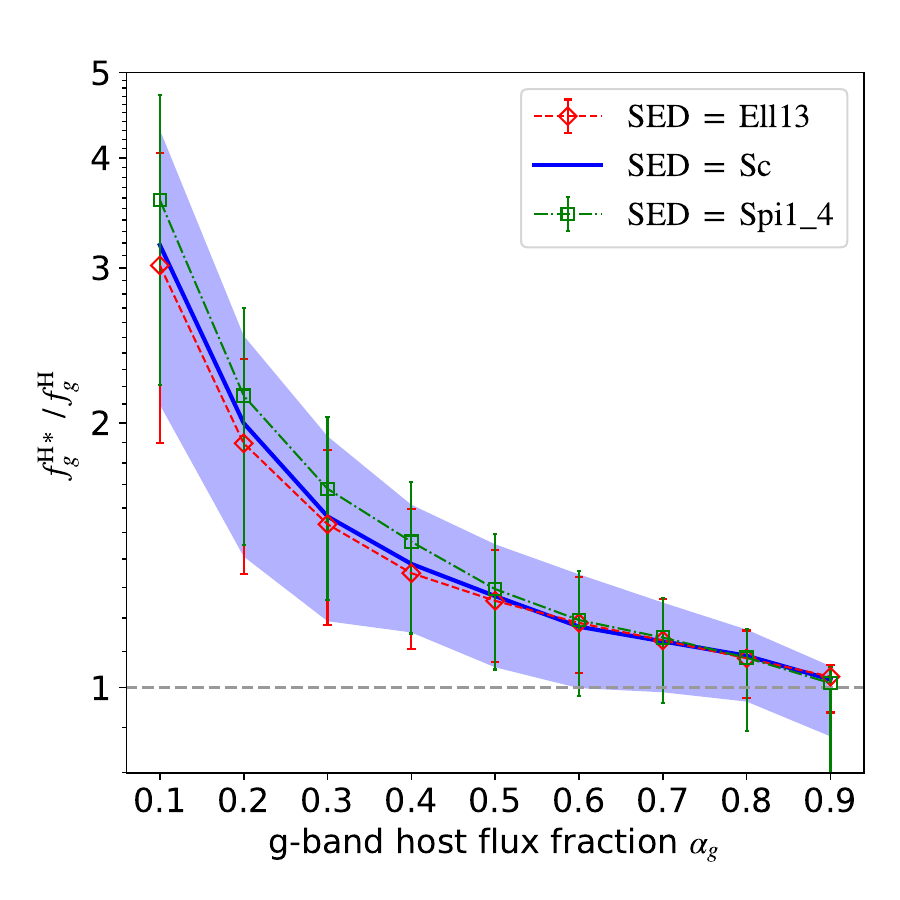}
    \end{minipage}
    \caption{The accuracy of the FVG method on retrieving the host galaxy flux, $f^{\rm H*}_g/f^{\rm H}_g$, as a function of the input $g$ band host galaxy flux fraction, $\alpha_{g}$, when changing the total time span (\textbf{upper left}), the observational cadence (\textbf{upper right}), the signal-to-noise ratio (\textbf{lower left}), and the assumed host galaxy SED (\textbf{lower right}). 
    The median (blue curve) and the 16th to 84th percentile range (violet region) of $f^{\rm H*}_g/f^{\rm H}_g$, inferred from 200 independent fiducial mocks with conditions of $T = 90$ days, $\Delta t = 3$ days, ${\rm S/N} = 50$, and the Sc SED, are compared to the median (symbols) and the 16th to 84th percentile range (error bars) inferred from other mocks with somewhat different conditions nominated in the legends.}
    \label{fig:change_one_para}
\end{figure}

\begin{figure}[H]

    \includegraphics[width=0.8\linewidth]{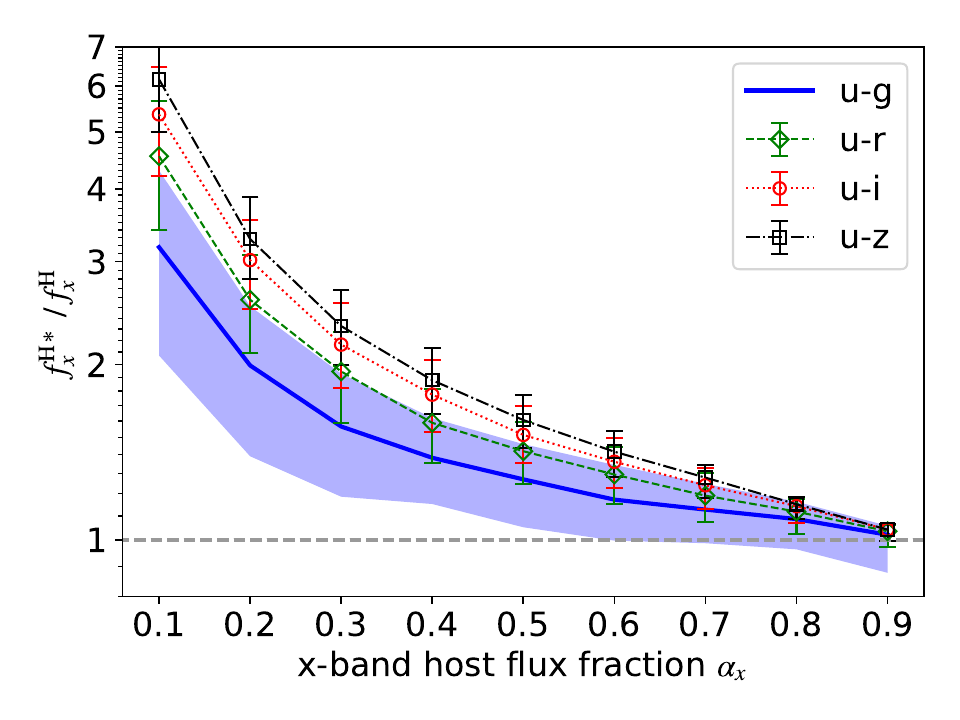}
    \caption{Analogous to Figure~\ref{fig:change_one_para} but for the median (blue curve or symbols) and the 16th to 84th percentile range (violet region or error bars) of $f^{\rm H*}_x/f^{\rm H}_x$ inferred from the fiducial mocks as a function of the input $x$ band host galaxy flux fraction, $\alpha_x$, when combining $u$ and $x$ bands, where $x$ is $g$, $r$, $i$, or~$z$.}
    \label{fig:how_FVG_overestimate_through_bands}
\end{figure}

We evaluate the accuracy of the FVG method for different input host galaxy flux fractions by comparing the medians and the 16th to 84th percentile ranges of $f^{\rm H\ast}_g / f^{\rm H}_g$ inferred from 200 independent simulations with different conditions (Section~\ref{sec:Mocking}). Figure~\ref{fig:change_one_para} clearly shows that the FVG method generally overestimates the host galaxy flux: more significant for the smaller intrinsic host galaxy flux fraction. Using $u$ and $g$ bands, the FVG method overestimates the host galaxy flux by a factor of $\sim$three when $\alpha_g = 0.1$.
Remarkably, this result remains unchanged for different observation durations, cadences, SNRs, and host galaxy types. 
For a shorter duration, $f^{\rm H\ast}_g / f^{\rm H}_g$ is slightly larger though with larger scatter.
Note that the large scatter of $f^{\rm H\ast}_g / f^{\rm H}_g$ is partially attributed to the random nature of AGN variability, especially for a shorter duration.
Furthermore, Figure~\ref{fig:how_FVG_overestimate_through_bands} confirms the above result by combining two different bands and highlights a larger overestimation when using two bands with a larger difference in wavelength. 
For $\alpha_g = 0.1$, the factor of overestimation by the FVG method increases to $\sim$six when combining $u$ and $z$ bands.

\section{Observational Tests}
\label{sec:Applications}

In Section~\ref{sec:simulation}, we found through simulations that the FVG method tends to overestimate the host galaxy flux. To further validate this conclusion, we compare the host galaxy fluxes inferred from the FVG method with those from different methods for three Seyfert galaxies: Mrk 509, Mrk 279, and 3C 120. Using multi-band light curves, these galaxies have been repeatedly used to demonstrate the validation of the FVG method, but the retrieved host galaxy fluxes are scarcely directly compared with other methods, such as image decomposition.

\subsection{Mrk 509}
\label{sec:Mrk509}

\begin{figure}[H]

    \includegraphics[width=0.8\linewidth]{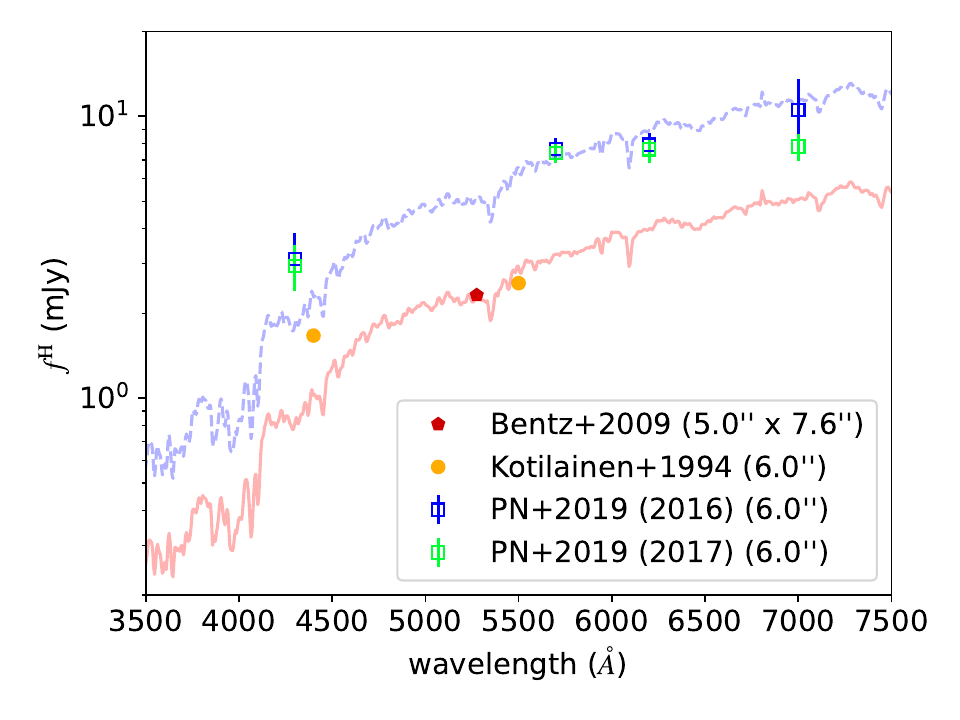}
    \caption{Comparison of the host galaxy fluxes of Mrk 509 reported by several studies. The open squares indicate the results obtained with the FVG method, while the other data points correspond to the results of the image decomposition method. The redshifted bulge galaxy template of~Kinney et~al.~\cite{SED_bulge} is fitted to the host galaxy fluxes obtained with the FVG (blue dashed curve) and the image decomposition (red solid curve) methods. \hl{The legend lists apertures applied in each study.} 
}
    \label{fig:Mrk509_hgfluxes}
\end{figure}

Mrk 509 is a bright nearby Seyfert 1 galaxy with redshift $z = 0.0344$~\citep{1993AJ....105.1637H} and has a bulge-type host galaxy revealed by Hubble Space Telescope (HST) imaging~\citep{2009ApJ...697..160B}. Mrk 509 is so close that its host galaxy flux can be measured by different methods. Using the image decomposing method,~\citet{1994MNRAS.266..953K} reported the $B$ band, $V$ band, $J$ band, and $K$ band host galaxy fluxes, while~\citet{2009ApJ...697..160B} reported the host galaxy flux at rest-frame 5500 $\angstrom$.
These results are presented in Figure~\ref{fig:Mrk509_hgfluxes}.

The host galaxy fluxes of Mrk 509 have also been estimated with the FVG method.~\citet{2019MNRAS.490.3936P} applied this approach to Mrk 509 using campaigns 2016 and 2017 and determined the host galaxy fluxes in four optical bands. These results are also depicted in Figure~\ref{fig:Mrk509_hgfluxes}.

Figure~\ref{fig:Mrk509_hgfluxes} clearly shows that the host galaxy fluxes inferred by the FVG method are larger than those obtained with the image decomposition. However, it is also necessary to examine whether the difference is caused by different aperture sizes used in these observations. While measuring the host galaxy fluxes of Mrk 509 with the FVG method, Pozo Nu{\~n}ez et~al. \cite{2019MNRAS.490.3936P} employed a 6.0$^{\prime\prime}$ aperture to minimize the absolute scatter for the fluxes. 
\citet{2009ApJ...697..160B} utilized three components (i.e., a Point Spread Function (PSF), a constant sky, and a host galaxy) to decompose the image captured by the HST and the MDM Observatory in an aperture of 5.0$^{\prime\prime}$ $\times$ 7.6$^{\prime\prime}$ and calculated the host galaxy fluxes after removing the PSF and sky components. 
\citet{1994MNRAS.266..953K} used three components (i.e., an AGN point source, a bulge, and an exponential disk) and utilized a 6$^{\prime\prime}$ aperture to determine the host galaxy fluxes after removing the AGN point source component. As we can see, there is no significant difference in the aperture size employed by them.

\subsection{Mrk 279}\label{sec:Mrk279}

\begin{figure}[H]
    \includegraphics[width=0.8\linewidth]{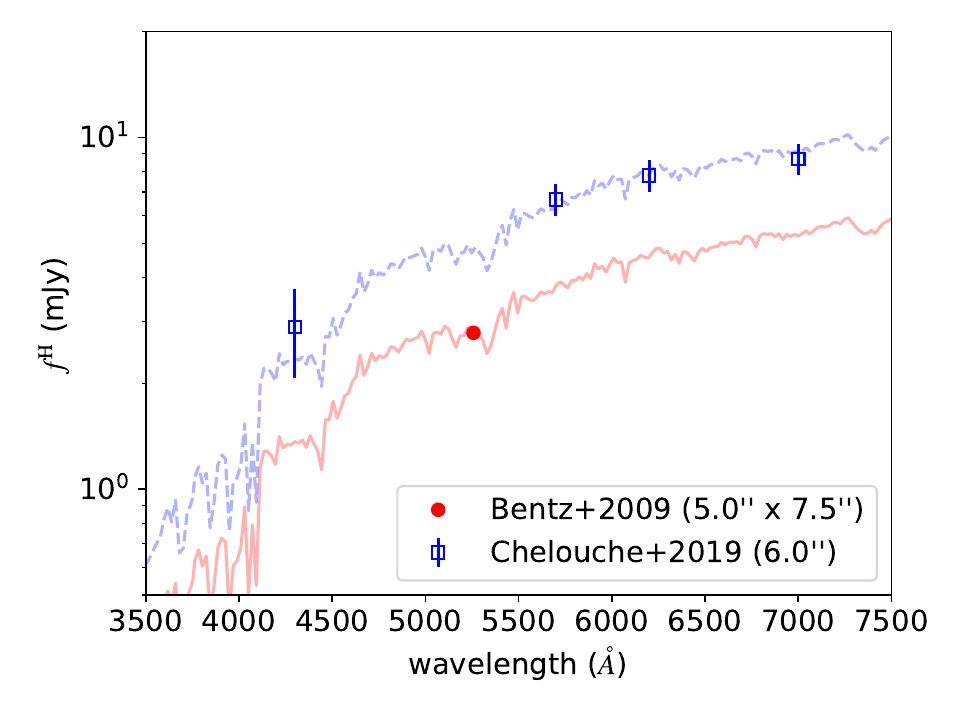}
    \caption{Same as Figure~\ref{fig:Mrk509_hgfluxes} but for Mrk 279 and a redshifted Sa galaxy template of~\citet{SED_swire}.}
    \label{fig:Mrk279_hgfluxes}
\end{figure}

Mrk 279, a nearby Seyfert 1 galaxy with redshift $z=0.031$, has an S0/Sa host galaxy in the high-resolution HST images~\citep{2002ApJ...569..624P}.~\citet{2009ApJ...697..160B} measured the host galaxy flux at rest-frame 5500 $\angstrom$ with the image decomposition method. More \mbox{recently,~\citet{2019NatAs...3..251C}} obtained the host galaxy fluxes in four optical bands with the FVG method. Their results are depicted in Figure~\ref{fig:Mrk279_hgfluxes}.
\vspace{-8pt}

\citet{2009ApJ...697..160B} employed an aperture of 5.0$^{\prime\prime}$ $\times$ 7.6$^{\prime\prime}$ and the same image decomposition method as Mrk 509. Meanwhile,~\citet{2019NatAs...3..251C} carried out photometry using a 6.0$^{\prime\prime}$ aperture and subsequently disentangled the host galaxy flux using the FVG method. Notably, the aperture size utilized by these two studies are similar.

\subsection{3C 120}\label{sec:3C120}

\begin{figure}[H]

   \hspace{-4pt} \includegraphics[width=0.8\linewidth]{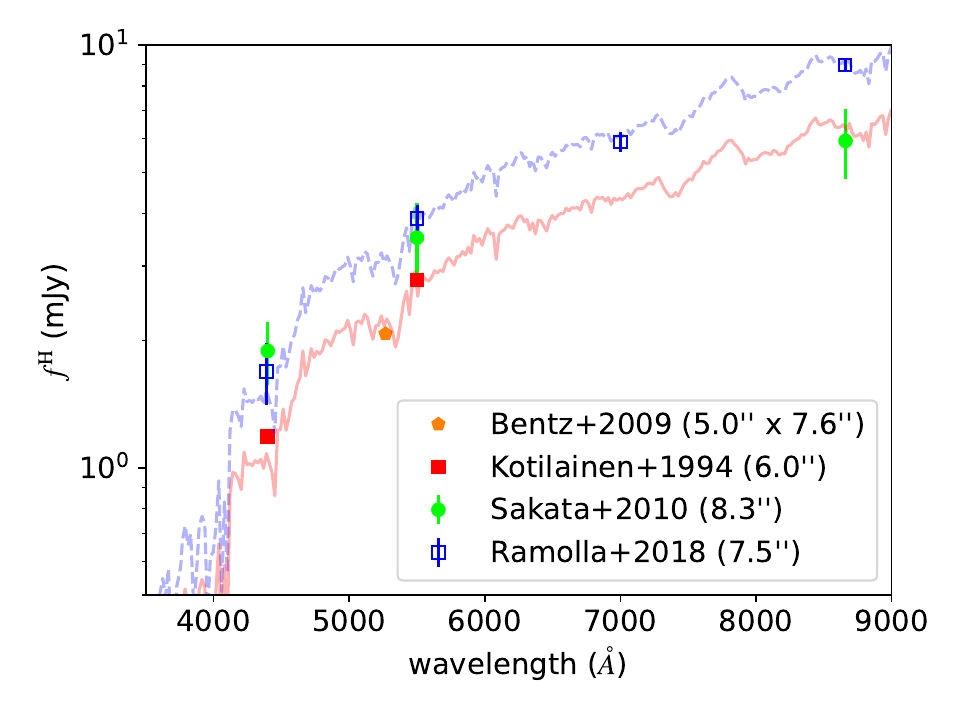}
    \caption{Same as Figure~\ref{fig:Mrk509_hgfluxes} but for 3C 120 and a redshifted S0 galaxy template of~\citet{SED_swire}.}
    \label{fig:3C120_hgfluxes}
\end{figure}

\textls[-15]{The host galaxy fluxes of 3C 120, a nearby radio galaxy located at a redshift of 0.033~\citep{1987ApJ...316..546W}, were measured in different bands with the image decomposition \mbox{method~\citep{1994MNRAS.266..953K, 2009ApJ...697..160B, 2010ApJ...711..461S}.} Ramolla et~al.~\cite{2015A&A...581A..93R} and~\citet{2018A&A...620A.137R} obtained the host galaxy fluxes in four optical bands by utilizing the FVG method. These results are shown in Figure~\ref{fig:3C120_hgfluxes}.}

\citet{2018A&A...620A.137R} utilized the FVG method with an aperture of 7.5$^{\prime\prime}$, while Bentz et~al.~\cite{2009ApJ...697..160B} employed the image decomposition method with an aperture of 5.0$^{\prime\prime}$~$\times$~7.6$^{\prime\prime}$ and obtained a smaller host galaxy flux. Furthermore,~\citet{1994MNRAS.266..953K} used a 6.0$^{\prime\prime}$ aperture and also obtained smaller host galaxy fluxes by using the image decomposition method. Even~\citet{2010ApJ...711..461S} adopted a larger aperture, i.e., 8.3$^{\prime\prime}$, than~\citet{2018A&A...620A.137R}, and their host galaxy flux at the longest wavelength obtained by the image decomposition method is still smaller than~\citet{2018A&A...620A.137R}.
\vspace{-6pt}

\section{Discussion}
\label{sec:discussion}

\subsection{The FVG Method Gives Rise to Larger Host Galaxy Fluxes than the Image Decomposition~Method}

Figures~\ref{fig:Mrk509_hgfluxes}--\ref{fig:3C120_hgfluxes} compare the host galaxy fluxes obtained by the FVG method and the image decomposition method for Mrk 509, Mrk 279, and 3C 120, respectively. Note that the host galaxy fluxes measured by these two methods are usually given at different wavelengths. To quantify to what extent the FVG method globally overestimates the host galaxy flux, we fit the assumed host galaxy SED of each source to the FVG-measured fluxes. We then vertically shift the assumed host galaxy SED to roughly align with the fluxes obtained by the image decomposition method since the uncertainties of these fluxes are mostly unavailable.
In this way, we crudely estimate that the FVG method tends to overestimate the host galaxy flux by a factor of $\sim$2.2, $\sim$1.7, and $\sim$1.4 for Mrk 509, Mrk 279, and 3C 120, respectively, which are summarized in Table~\ref{tab:overestimation_of_FVG}.

\begin{table}[H] 
\caption{Overestimation of the host galaxy flux for each object.\label{tab:overestimation_of_FVG}}

\begin{adjustwidth}{-\extralength}{0cm}
\begin{tabularx}{\fulllength}{Cc}
\toprule
\textbf{Object}	& \textbf{Overestimated Factor of the FVG Method Relative to the Image Decomposition Method}\\
\midrule
Mrk 509		& 2.237\\
Mrk 279		& 1.722\\
3C 120      & 1.410\\
\bottomrule
\end{tabularx}
\end{adjustwidth}
\end{table}

The validity of the FVG method has generally been demonstrated by simply comparing the shape of the FVG-measured fluxes to that of the assumed host galaxy SED~\citep{2018A&A...620A.137R, 2019NatAs...3..251C, 2019MNRAS.490.3936P}, rather than directly comparing the FVG-measured fluxes to those inferred by other methods, such as image decomposition. 
Figures~\ref{fig:Mrk509_hgfluxes}--\ref{fig:3C120_hgfluxes} confirm that the shape of the FVG-measured fluxes is consistent with that of the assumed host galaxy SED. However, the FVG-measured fluxes are almost larger than those obtained by the image decomposition method. 
This suggests that solely comparing the shape of the FVG-measured fluxes to that of the assumed host galaxy SED is not enough in justifying the FVG method.

\subsection{Comparison with Previous Works}

Some studies have compared the host galaxy fluxes obtained by the FVG method with results from other methods. Compared to the image decomposition method, Haas et~al.~\cite{2011A&A...535A..73H} obtained higher host galaxy fluxes for PG0003+199 using the FVG method. They suggest that such overestimation may not be solely attributed to the failure of the FVG method but may be related to inaccuracies in the fitting procedure of GALFIT and the image decomposition software they used. They argue that if the host galaxy flux obtained by GALFIT is accurate, the AGN variability curve at the faint end of the flux--flux plot would show a strong curvature towards redder colors, contradicting the findings of~Sakata et~al.~\cite{2010ApJ...711..461S}. However, in~\citet{2010ApJ...711..461S}, what falls on the linear extension of the AGN slope is the sum of the host galaxy flux and the narrow-line region flux, not just the host galaxy flux alone. Considering only the host galaxy flux, it would still fall on the left side of the AGN slope, consistent with our conclusion obtained in Section~\ref{sec:simulation}.

When~\citet{2012A&A...545A..84P} studied 3C 120, they compared the host galaxy fluxes derived from the FVG method with those obtained by~\citet{2006ApJ...644..133B} and~Bentz et~al.~\cite{2009ApJ...697..160B} using GALFIT. They found that their results were located between those of Bentz et~al.~\cite{2006ApJ...644..133B} and~\citet{2009ApJ...697..160B}. 
However, due to the improvements made by~\citet{2009ApJ...697..160B} over~\citet{2006ApJ...644..133B} in refining the model for fitting the images and allowing for better matching with observational data, {\citet{2009ApJ...697..160B}} obtained more accurate host galaxy fluxes. If the results obtained by~\citet{2012A&A...545A..84P} are directly compared with those of~\citet{2009ApJ...697..160B}, it would be found that the former significantly exceeds the latter. This is again in agreement with our conclusion.

\citet{2019ApJ...886..150M} compared the host galaxy fluxes for 25 quasars at $z<0.6$ using spectral decomposition, image decomposition, and FVG methods. It was found that the results obtained using the FVG method were significantly higher than those obtained from the other two methods, with larger errors. Such findings align with our results.

\subsection{A Likely Invalid Assumption for the FVG Method}

On the nature of the bluer-when-brighter pattern of AGN, the first explanation put forth by~\citet{1981AcA....31..293C} attributes this pattern to a combination of the stable host galaxy and the variable AGN with a constant spectral shape. This assumption has long been the foundation for the FVG method.
However,~\citet{2014ApJ...783..105R} suggested that the pattern can be better explained by a simple inhomogeneous disk model that features large localized temperature fluctuations. 
Moreover, as demonstrated by~\citet{2014ApJ...792...54S}, a significant bluer-when-brighter trend is observed in quasars where the host galaxy emission is negligible and so the bluer-when-brighter trend is intrinsic to the AGN accretion physics.
Additionally, the~\citet{1981AcA....31..293C} assumption that the AGN component has a constant color was suggested to be inadequate in the ultraviolet band by~\citet{2011ApJ...731...50S}. 
Based on these considerations, it can be concluded that the hypothesis of ``stable host galaxy plus variable AGN'' is not a viable explanation for the bluer-when-brighter trend of AGN.

\subsection{Can Involving More than Two Photometric Bands in the FVG Method Help Alleviate the Overestimation?}

{Since the traditional FVG method operates on a flux--flux plot defined by a pair of two bands, most previous studies using the FVG method (as we cited in Figures~\ref{fig:Mrk509_hgfluxes}--\ref{fig:3C120_hgfluxes} for three AGN) only used two photometric bands and repeated the analysis for each pair of two bands. To fully make use of multi-band information,~\citet{2022A&A...657A.126G} proposed an upgraded probabilistic FVG method (as mentioned in Section~\ref{sec:intro}), which considers all photometric bands simultaneously. They found that when using the probabilistic FVG method, the retrieved host galaxy fluxes of Mrk 509, Mrk 279, and 3C 120 are similar to (though with smaller uncertainties than) those obtained using the traditional FVG method (see Figures 8 to 10 in~\citet{2022A&A...657A.126G}). Therefore, considering more than two photometric bands does not help alleviate the overestimation of the FVG method.
}


\subsection{Improving the Thermal Fluctuation Model Adopted}\label{sect:tfm}

{The adopted thermal fluctuation model~\citep{Cai2018,Cai2020} can not only account for the timescale-dependent color variation (i.e., more significant bluer-when-brighter at shorter timescales) of NGC 5548 unveiled by~\citet{Zhu2018ApJ...860...29Z} but also reproduce the inter-band lag and correlation across optical/UV to X-ray in four local Seyfert galaxies well~\citep{Cai2020}. Two of them have puzzling large UV-to-X-ray lags. 
Both the timescale-dependent color variation and the puzzling large UV-to-X-ray lags are challenges to the widely accepted reprocessing model. 
Thanks to the success of the thermal fluctuation model, it is worthy of further development in two aspects. 
On one hand, there are many observed properties of AGN, such as the broadband SED across optical/UV to X-ray, which should be simultaneously and self-consistently explained together with the properties of variability explored \mbox{by~\citet{Cai2018,Cai2020}.}
On the one hand, a new thermal fluctuation model involving disk wind or outflow is attractive as suggested by the prevalent winds likely required to address the universal average SED for quasars from the optical to the extreme UV~\citep{Cai2023NatAs...7.1506C}.
In addition, utilizing other models for AGN variability such as~\citep{Sun2020ApJ...891..178S,Kammoun2024A&A...686A..69K} would be deserved in future works to justify the FVG method.}

\section{Conclusions}
\label{sec:conclusion}

We simulated the AGN light curves using the thermal fluctuation model and mocked the total (AGN + host galaxy) light curves assuming a series of host galaxy fractions and under different observational conditions. By applying the FVG method to retrieve the input host galaxy flux, we observed that 
the FVG method nearly always overestimates the host galaxy contribution, regardless of the observational duration, cadence, and SNR.

Furthermore, considering three bright local Seyfert galaxies, namely Mrk 509, Mrk 279, and 3C 120, whose multi-band variability has been used to demonstrate the validity of the FVG method by previous works~\citep{2018A&A...620A.137R, 2019NatAs...3..251C, 2019MNRAS.490.3936P}, we showed that the FVG method tends to overestimate the host galaxy flux more than the image decomposing method. This is in agreement with~\citet{2019ApJ...886..150M}, who compared the host galaxy fluxes of 25 quasars obtained by three different methods. The overestimation of the FVG method is not attributed to the different aperture sizes involved by the data available for different methods.

Our study suggests that the FVG method could only be relatively reliable for a moderate host galaxy fraction, 
but it is not suitable for quasars whose total fluxes are dominated by AGN. Thus, we caution that applying the FVG method should be taken carefully in the era of time-domain astronomy~\citep{Brandt2018arXiv181106542B,Wang2023SCPMA..6609512W}.

\vspace{6pt} 

\authorcontributions{Conceptualization, Z.-Y.C.; methodology, M.-X.C., Z.W., and Z.-Y.C.; validation, Z.-Y.C., Z.W., L.-L.F., and J.-X.W. All authors have read and agreed to the published version of the manuscript.}

\funding{\hl{This research was funded by the National Key Research and Development Program of China, grant number 2023YFA1608100; the Strategic Priority Research Program of the Chinese Academy of Sciences, grant number XDB 41000000; the National Science Foundation of China, grant numbers 12373016 and 12033006; the National Natural Science Foundation of China, grant numbers 12173037 and 12233008; the CAS Project for Young Scientists in Basic Research, grant number YSBR-092; the China Manned Space Project, grant numbers CMS-CSST-2021-A04 and CMS-CSST-2021-A06; the Fundamental Research Funds for the Central Universities, grant number WK3440000006; and the Cyrus Chun Ying Tang Foundations.} 
}

\dataavailability{The data underlying this article will be shared upon reasonable request to the corresponding author.}

\acknowledgments{\hl{We thank Zhen-Bo Su and Biao Zhang for the useful discussions. We are grateful for the helpful and constructive feedback provided by the referees.} 
}

\conflictsofinterest{The authors declare no conflicts of interest.}

\abbreviations{Abbreviations}{
The following abbreviations are used in this manuscript:\\

\noindent 
\begin{tabular}{@{}ll}
AGN & Active galaxy nuclei\\
FVG & Flux variation gradient\\
SDSS & Sloan Digital Sky Survey\\
SED & Spectral energy distribution\\
SNR & Signal-to-noise ratio\\
MCMC & Markov Chain Monte Carlo\\
HST & Hubble Space Telescope\\
PSF & Point Spread Function\\
UV & Ultraviolet
\end{tabular}
}

\begin{adjustwidth}{-\extralength}{0cm}

\reftitle{References}

\PublishersNote{}
\end{adjustwidth}
\end{document}